\documentclass[aps,prl,preprint,superscriptaddress]{revtex4}

\usepackage{units}
\usepackage[dvips]{graphicx}

\begin{document} 

\title{Magnetic Anisotropy and Magnetization Dynamics of Individual Atoms and Clusters of Fe and
  Co on Pt(111)}

\author{T. Balashov}
\affiliation{Physikalisches Institut, Universit\"at Karlsruhe (TH), 76131 Karlsruhe, Germany}

\author{T. Schuh}
\affiliation{Physikalisches Institut, Universit\"at Karlsruhe (TH), 76131 Karlsruhe, Germany}

\author{A. F. Tak\'acs}
\affiliation{Physikalisches Institut, Universit\"at Karlsruhe (TH), 76131 Karlsruhe, Germany}

\author{A. Ernst}
\affiliation{Max-Planck-Institut f\"ur Mikrostrukturphysik, 06120 Halle, Germany}

\author{S. Ostanin}
\affiliation{Max-Planck-Institut f\"ur Mikrostrukturphysik, 06120 Halle, Germany}

\author{J. Henk}
\affiliation{Max-Planck-Institut f\"ur Mikrostrukturphysik, 06120 Halle, Germany}

\author{I. Mertig}
\affiliation{Max-Planck-Institut f\"ur Mikrostrukturphysik, 06120 Halle, Germany}
\affiliation{Martin-Luther-Universit\"at Halle-Wittenberg, Institut
  f\"ur Physik, 06099 Halle, Germany}

\author{P. Bruno}
\affiliation{Max-Planck-Institut f\"ur Mikrostrukturphysik, 06120 Halle, Germany}
\affiliation{European Synchrotron Radiation Facility, 38043 Grenoble Cedex, France}

\author{T. Miyamachi}
\affiliation{Graduate School of Engineering Science, Osaka University, Japan}

\author{S. Suga}
\affiliation{Graduate School of Engineering Science, Osaka University, Japan}

\author{W. Wulfhekel}
\affiliation{Physikalisches Institut, Universit\"at Karlsruhe (TH), 76131 Karlsruhe, Germany}

\date{\today}

\begin{abstract}
  The recently discovered giant magnetic anisotropy of single magnetic
  Co atoms raises the hope of magnetic storage in small clusters.  We
  present a joint experimental and theoretical study of the magnetic
  anisotropy and the spin dynamics of Fe and Co atoms, dimers, and trimers
  on Pt(111). Giant anisotropies of individual atoms and clusters as
  well as lifetimes of the excited states were determined with
  inelastic scanning tunneling spectroscopy. The short lifetimes due
  to hybridization-induced electron-electron scattering oppose the
  magnetic stability provided by the magnetic anisotropies.
\end{abstract}

\maketitle 

In modern magnetic recording, bits are stored in magnetically stable
metallic grains, with larger densities being realized by reduced grain
sizes. The magnetic stability is related to the grain's magnetic
anisotropy energy (MAE) that has to be overcome to reverse the
magnetization. Recently, a giant MAE of Co atoms
on Pt(111) of \unit[9.3]{meV} was found with X-ray magnetic circular
dichroism (XMCD) \cite{Gambardella03}. This raised the hope for
achieving the ultimate size limit in magnetically stable atoms or
clusters at cryogenic temperatures. The challenge of studying magnetic
stability lies in the direct investigation of a single atom, which is
not possible by XMCD\@. We report here on the MAE and magnetization
dynamics of single Fe and Co atoms and clusters on Pt(111) investigated
with scanning tunneling microscopy (STM) and show that the
quantum-mechanical nature of the magnetic system becomes essential.

To determine the MAE and the magnetization dynamics of individual
atoms and small clusters, we performed inelastic tunneling
spectroscopy (ITS) with a home-built low-temperature STM
in ultra-high vacuum. The tunneling electrons may exchange spin
angular momentum with these magnetic objects \cite{Heinrich04}. 
The result of such inelastic spin-flip scattering is a change of the
magnetization direction of the object \cite{Balashov08}.  For Fe and
Co atoms and clusters on Pt(111) the uniaxial anisotropy dominates and favors
perpendicular anisotropy \cite{Gambardella03,Etz08}.  In the quantum
limit, i.\,e.\ for an isolated magnetic object with spin $S$, the uniaxial
anisotropy energy can be written as $D S_{z}^{2}$ ($D < 0$ for easy axis out of plane).  This
description is linked to the classical uniaxial MAE $K \cos^{2}\theta$
by the correspondence principle $\cos\theta = S_{z} / S$ ($\theta$~is
the angle of the magnetization with respect to the surface normal). To
relate the spin-flip energy $E_{\mathrm{sf}}$ of a tunneling electron
to the MAE, we note that upon spin-flip scattering the magnetic
cluster is excited from its ground state with $S_{z} = \pm S$ to a
state with $S_{z} = \pm (S - 1)$, i.\,e.\ $S_{z}$ of the object is
changed by $1$.  For known $S$, $K$ can be estimated as
$E_{\mathrm{sf}} \times S^{2} / (2 S - 1)$ and $|D| (2 S - 1) =
E_{\mathrm{sf}}$.

Inelastic spin-flip scattering shows up in ITS\@.  This approach was
so far restricted to atoms on insulating layers, e.\,g.\ Mn atoms on
Al$_2$O$_3$ or CuN~\cite{Hirjibehedin07,Heinrich04}
but failed in entirely metallic systems. Recently, we have shown that even
in metallic structures (bulk samples and thin films) spin-flip
scattering can be detected \cite{Balashov08,Balashov06,Gao08}.  Here
we extend this technique to single atoms and clusters on Pt(111).  In
metallic systems, elastic tunneling leads to a linear dependence of
the tunnel current $I$ on the bias voltage $U$ in the low bias regime.
If $\mathrm{e}\,U = E_{\mathrm{sf}}$, an additional inelastic tunnel
channel opens and the slope of $I(U)$ is increased for $\mathrm{e}\,U >
E_{\mathrm{sf}}$. This change is usually too small to be identified
directly. It can, however, be seen as a peak in $\mathrm{d}^{2} I /
\mathrm{d}U^{2}$ \cite{Wolf85,Stipe98}.  The excitation occurs for
both tunneling directions, with a `negative peak' appearing at negative
bias.

\begin{figure}
  \centering
  \includegraphics[width=0.9\columnwidth,angle=0,clip]{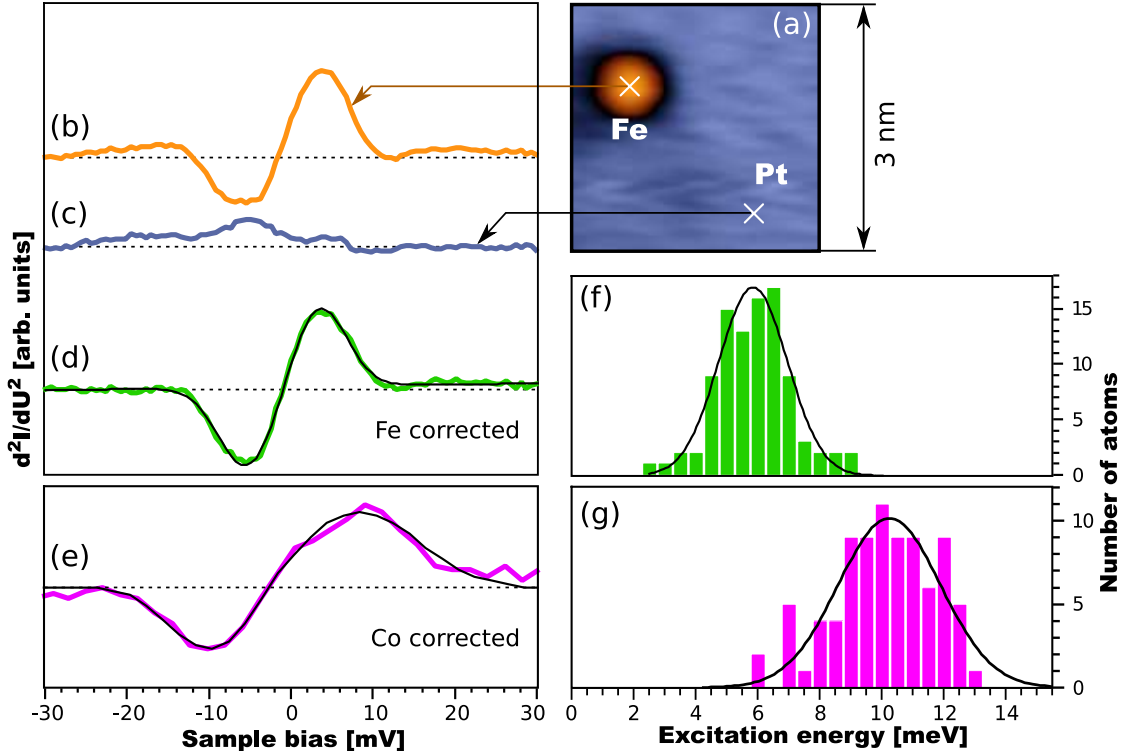}
  \caption{(a) Topography of a single Fe atom on Pt(111). 
    (b) $\mathrm{d}^{2} I / \mathrm{d} U^{2}$ spectra of an
    Fe atom on Pt(111), (c) of the bare Pt(111) background, and
    (d) the difference spectrum. (e) Background-corrected Co spectrum. Gaussian fits are shown in black. 
    (f) Distributions of excitation energies for Fe atoms
    and (g) for Co atoms.}
  \label{first_peak}
\end{figure}

To achieve a high surface quality, the Pt(111) substrate was cleaned
by multiple cycles of Ar sputtering and annealing until contaminations
were absent in STM images.  W tips were cleaned in vacuum by flashing,
such that the end of the tip melted.  Small amounts ($< 0.01$ atomic
layer) of Fe or Co were deposited on Pt(111) at $\unit[4.3]{K}$,
resulting in isolated adatoms.  $\mathrm{d}^{2} I / \mathrm{d} U^{2}$
spectra were recorded for Fe atoms and the Pt surface (crosses in
Fig.~\ref{first_peak}a). The second derivative was measured
with a lock-in amplifier detecting the second
harmonics in the tunnel current due to a $\unit[1.4-3.2]{mV}$,
$\unit[16.4]{kHz}$ modulated bias voltage.  While spectrum (b) of the
Fe atom on Pt(111) (see Fig.~1) clearly shows an `inelastic' minimum/maximum
structure, the Pt spectrum (c) displays a minute signal. The genuine
excitation spectrum (d) is obtained by subtracting (c) from (b) and is
found to be almost symmetric, with the minimum and the maximum
reflecting inelastic excitations. An excitation energy of $\approx
\unit[6]{meV}$ is estimated from a Gaussian fit of peak and dip.  Co
atoms on Pt(111) give very similar results, providing $\approx \unit[10]{meV}$
(see~Fig.~\ref{first_peak}e).

The observed spectroscopic features are ascribed to spin-flip excitations
since other possibilities can be excluded as follows.
(i) Plasmons in Fe, Co and Pt have energies in the range of several \unit{eV}.
Besides, they cannot be confined to single atoms, and therefore do not
explain the observed excitations.
(ii) Atomic vibrations can be excluded by first-principles calculations.
The vibration energies were obtained with the Vienna Ab initio Simulation
Package (VASP) \cite{Kresse96}.  The potential landscape was mapped by
moving the adatom from its equilibrium position in all directions.
Within the harmonic approximation, the softest phonon was estimated to
\unit[27]{meV} for Fe and \unit[24]{meV} for Co.  These energies are
significantly larger than the measured excitation energies.
(iii) The Kondo effect could produce
similar features in the spectra \cite{Kondodip}.  However, Fe and Co
on Pt(111) show no Kondo effect: The perpendicular magnetic anisotropy
of Co and Fe on Pt(111) \cite{Gambardella03,Etz08} lifts the spin
degeneracy of the atom and hinders the exchange of electrons with
opposite spins between the atom and the substrate
\cite{Leuenberger06,Otte08}.  Further, being easily polarized, the
magnetic moments at the Pt atoms adjacent to the adatom
\cite{Gambardella03,Herrmannsdoerfer96,Etz08} counteract the Kondo
screening. Thus, the observed inelastic features are clearly
attributed to magnetic excitations.

To determine the classical MAE from the excitation energy, the spin
$S$ of the adsorbed atom and the induced Pt moments must be known.  We
therefore calculated the latter from first principles within our
scalar-relativistic Korringa-Kohn-Rostoker (KKR) approach. The large
polarizability of Pt demands to treat an `extended impurity' which
comprises the adatom and 5 Pt layers with 16 Pt atoms each. The
impurity is embedded in semi-infinite Pt(111) by solving the Dyson
equation for the Green function.~\cite{Wildberger95} Exact positions
of the Fe and Co adatoms, which are crucial for calculating reliable
electronic and magnetic properties, are obtained by
VASP\@. \cite{PhysRevB.67.125412,conte:014416}. The adatoms are attracted by
the surface but the inward relaxation does not strongly depend on the
adsorption site (fcc of hcp).  The adatom-tip interaction was 
simulated and found to reduce the inward relaxation by
\unit[2--3]{\%}.

\begin{figure}
  \centering
  \includegraphics[width = 0.6\columnwidth]{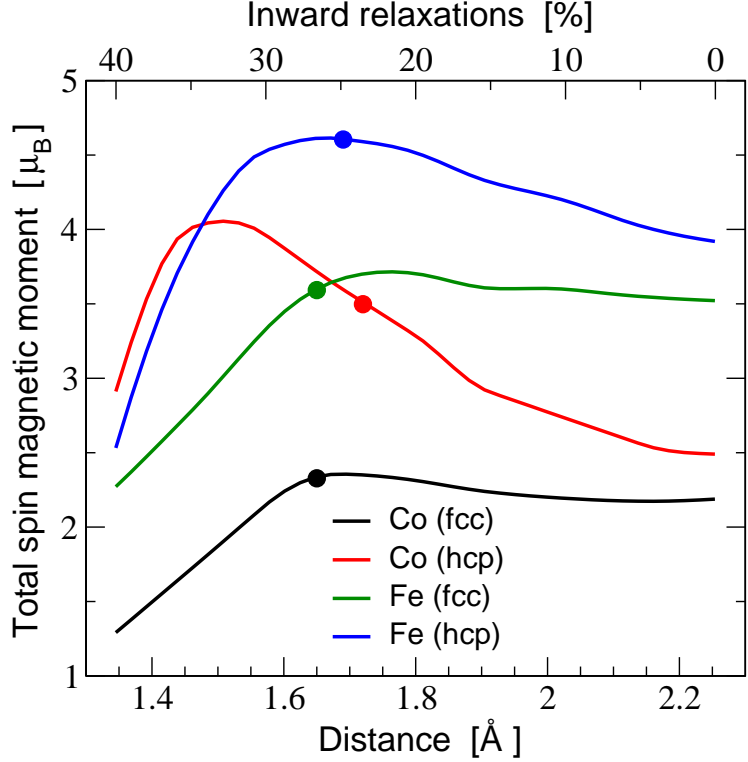}
  \caption{Spin magnetic moments versus relaxation of the Fe and Co
    adatoms. The circles indicate the equilibrium distances.}
  \label{fig:theory}
\end{figure}

The total spin magnetic moments (adatom and adjacent Pt atoms) show a
maximum at \unit[25--30]{\%} of inward relaxation and decrease with
increasing atom--substrate distance (see Fig.~\ref{fig:theory}). This
originates mainly from the induced Pt magnetic moments while the local
magnetic moments of Co and Fe are not affected by relaxation. Assuming
bulk positions for the adatoms results in significantly larger spin
moments \cite{Gambardella03,Etz08}.

From the total spin magnetic moment of \unit[2.2]{$\mu_{\mathrm{B}}$} for a
Co adatom in an fcc position we deduce $S = 1$, assuming a Land\'e
factor of $g \approx 2$. As a consequence, the excitation energy from
the ground state with $S_{z} = 1$ to the excited state with $S_{z} =
0$ corresponds directly to the classical uniaxial MAE \cite{footnote}.

The excitation energies were measured repeatedly
(see Fig.~\ref{first_peak}g). The distribution fitted with a Gaussian
yields an average MAE of $\unit[10.25]{meV}$ per Co atom
(cf. Table~\ref{tab:1}).  The result agrees well with the experimental
observations on an ensemble of Co atoms using XMCD
\cite{Gambardella03}, thereby confirming our conclusion that the
excitation is of magnetic origin. A similar analysis was performed
for Fe adatoms. The total spin moment of \unit[3.4]{$\mu_{\mathrm{B}}$}
for fcc sites results in $S =3/2$. The Gaussian fit in Fig.~\ref{first_peak}f
gives an average MAE of \unit[6.53]{meV} per Fe atom (see Table~\ref{tab:1}).

\begin{table}
  \caption{Measured excitation energies related to the MAE\@.}
  \begin{ruledtabular}
  \begin{tabular}{cccc}
      cluster & \begin{tabular}{c}spin\\transition\end{tabular} & \begin{tabular}{c}excitation\\energy [meV]\end{tabular} &  MAE [meV/atom] \\ \hline
      Co$_1$ & 1$\rightarrow$0 & 10.25$\pm$0.15 & 10.25$\pm$0.15\\
      Co$_2$ & 2$\rightarrow$1 & 8.2$\pm$0.4 &  5.5$\pm$0.3\\ 
      Co$_3$ & 3$\rightarrow$2 & 8.3$\pm$1.2 &  5.0$\pm$0.8\\ \hline
      Fe$_1$ & 3/2$\rightarrow$1/2 & 5.83$\pm$0.08 &  6.53$\pm$0.09\\
      Fe$_2$ & 3$\rightarrow$2 & 5.98$\pm$0.09 &  5.20$\pm$0.09\\ 
      Fe$_3$ & 9/2$\rightarrow$7/2 & 6.5$\pm$0.2 &  5.5$\pm$0.2
    \end{tabular}
  \end{ruledtabular}
  \label{tab:1}
\end{table}

Single adatoms can occupy fcc or hcp threefold hollow sites on
Pt(111). The potential barrier of $\approx\unit[0.2]{eV}$
\cite{Ternes08} for hopping from an fcc to an hcp site hinders thermal
diffusion between the two positions at \unit[4.3]{K}, suggesting that
adatoms occupy both sites equally likely. Interestingly, the
histograms for Fe and Co atoms are fitted well with a single Gaussian.
This observation implies that either the excitation energies do not
vary with position (within the resolution of $\approx \unit[3]{meV}$)
or that only one of the two positions is occupied.

To corroborate the experiment we computed MAE's using our relativistic
layer-KKR code, invoking the magnetic force theorem \cite{Wang96b}.
The MAE of a relaxed Fe atom is approximately \unit[3.2]{meV} in fcc and
\unit[0.4]{meV} in hcp positions; for a Co atom we find \unit[3.1]{meV}
for fcc and \unit[3.8]{meV} for hcp positions. These values are smaller than the
experimental ones but of the same order of magnitude.  The MAE for Fe
in hcp position is significantly smaller than the experimental value,
suggesting that Fe in the fcc position is mainly probed: Due to large
tunnel currents, the adatoms are possibly moved to the more stable fcc
sites.  In case of Co, there is no significant difference in MAE on
the hcp and fcc sites.  The agreement with experiment is substantially
better for Fe than for Co, pointing to the importance of correlation
effects in systems with localized $3d$ electrons. In the present
calculations, correlations are treated within the local spin density
approximation (LSDA) which works better for Fe than for Co:
The additional $d$ electron (with respect to Fe) causes stronger
correlations and requires to go beyond LSDA\@.

The probability of spin-flip scattering, as determined from the area
under the inelastic peaks, is $\approx\unit[2]{\%}$ for both Fe and Co
if the bias energy exceeds the MAE\@. This implies that at high bias
even with tunnel currents in the $\unit{pA}$ range, the magnetic state
of the atoms is flipped at a rate too high to be observed in STM. This
explains the absence of a hysteresis for single adatoms recently found
at a sample bias of $\approx\unit[200]{mV}$ \cite{Meier08}. Note also
that the peaks in the spectra are rather broad, an effect
accounted for by short lifetimes but not by thermal smearing or the
lock-in modulation. Having corrected for experimental broadening
\cite{Stipe98}, the widths of the inelastic peaks of Fe and Co are
$\unit[5.6]{meV}$ and $\unit[18]{meV}$, respectively, which correspond
to lifetimes of the excited states of less than $\unit[60]{fs}$ for Fe
and $\unit[24]{fs}$ for Co atoms. These lifetimes are shorter than the
time between consecutive spin-flip events ($\approx \unit[0.5]{ns}$ at
$I \approx \unit[10]{nA}$ with \unit[2]{\%} scattering probability).
Therefore only excitations from the ground state are observed.

Lifetimes provide information on the magnetization dynamics which is
not accessible by XMCD\@. The above lifetimes indicate efficient
relaxation processes which are absent for magnetic ions in insulators
\cite{Hirjibehedin07}.  Since comparable MAE's were found in metals
and insulators, spin-orbit interaction can be ruled out as relaxation
mechanism.  More likely, the strong hybridization of the adatom states
with those of the Pt substrate leads to efficient electron-electron
scattering processes that relax the magnetic state of the adatom: The
smaller hybridization of the Fe-$3d$ states with the Pt-$5d$ states as
compared to Co leads to both a lower MAE and to a longer lifetime of
the excited Fe state than that of Co.

\begin{figure}
  \centering
  \includegraphics[width=0.8\columnwidth,angle=0,clip]{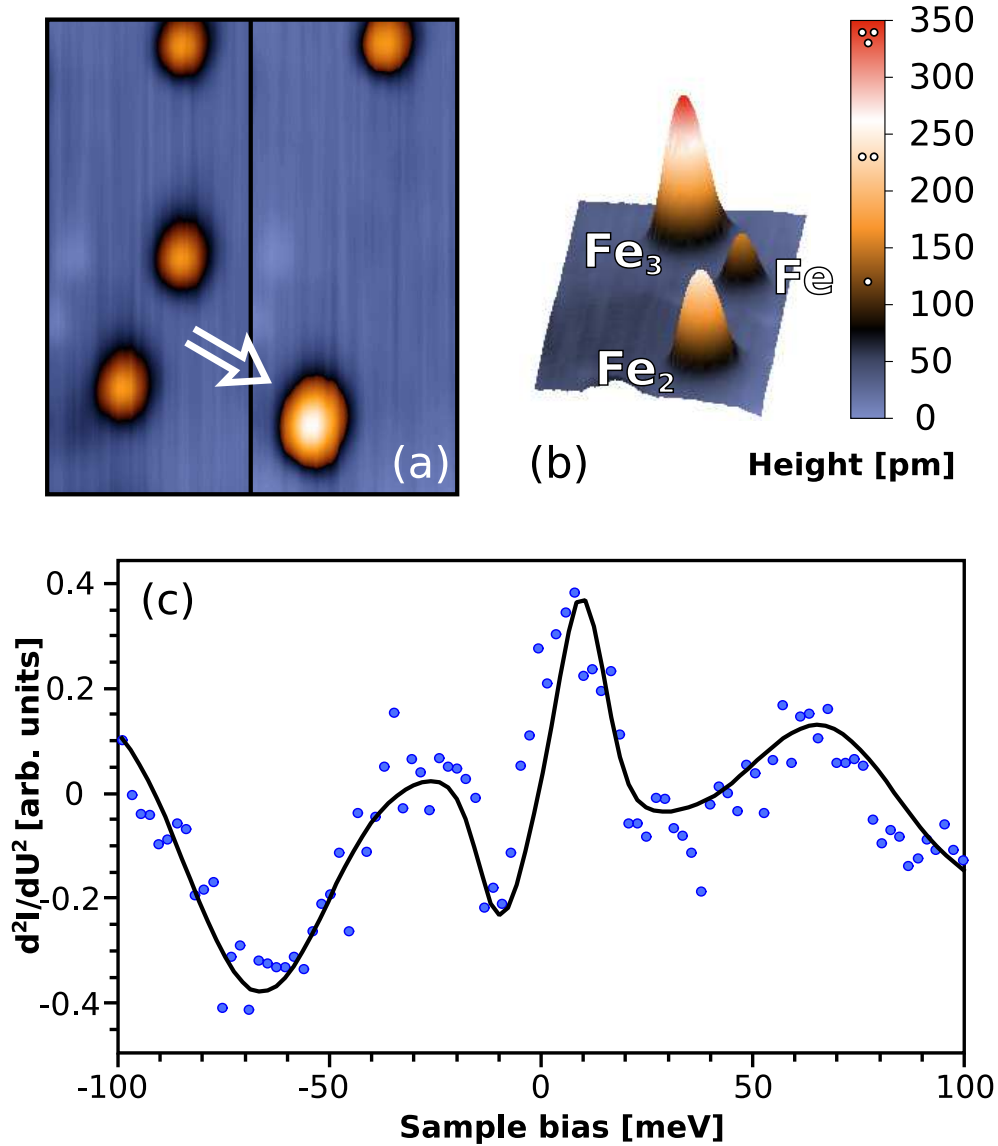}
  \caption{(a) Formation of a Co dimer by manipulating two Co atoms by the STM [arrow].
    (b) 3D view of an Fe atom, dimer and trimer. (c) $\mathrm{d}^{2}I / \mathrm{d}U^{2}$ spectra of an Fe
    dimer on Pt(111) with two excitations visible. Gaussian fits
    are shown as solid line.}
  \label{topography}
  \label{dimer} 
\end{figure}

Using the atomic manipulation capabilities of the STM \cite{Eigler90},
it is possible to extend the investigation to well-defined Fe and Co
clusters composed of two and more atoms (here: dimers and trimers;
Fig.~\ref{topography}a).  The dimers---and even more trimers---appear
significantly higher than single atoms in topographic scans
(Fig.~\ref{topography}b; cf.\ Ref.~\onlinecite{Chen99}).

To investigate the inelastic excitations of clusters we use the same approach
as for single atoms. In Fe and Co clusters the local magnetic moments
couple ferromagnetically, and the total spin of the cluster is obtained
by adding individual spins. There are, however, two different types
of spin-flip excitations in clusters, as illustrated by a low- and a high-energy
excitations in the dimer spectrum in Fig.~\ref{dimer}c.
(i) At low energies, the magnetic moment of the entire cluster can be
rotated collinearly, i.\,e.\ the total spin $S$ of the cluster is rotated
but its length stays constant. In analogy to single atoms
we obtain the MAE of the cluster from this excitation energy
(see Table~\ref{tab:1}). In agreement with XMCD data and calculations
\cite{Gambardella03}, the MAE per atom drops with the size of the
cluster. This is due to quenching of the orbital moment.  While a
single adatom on Pt(111) has a high rotational symmetry with respect
to rotation about the surface normal, this symmetry is broken with
adding another atom. The third atom lowers the MAE only slightly.  The
experiment shows that such a tendency does not depend on the adatom
species. The MAE's of Co trimers agree as well with XMCD data.

(ii) At higher energies, the `inner' magnetic structure of the cluster can be
excited into a noncollinear state. In addition to the anisotropy
energy, exchange energy has to be paid, being typically one order of
magnitude larger.
High energy excitations appear in experiment as
broad peaks around $\unit[50]{meV}$ for Fe dimers (Fig.~\ref{dimer}c).
The width of $\unit[33]{meV}$ is caused by the short lifetime of the
excited state ($\approx \unit[10]{fs}$). An evaluation of 18 dimers
gives an excitation energy of $\unit[54 \pm 2]{meV}$.  Assuming a spin
of $S = 3$ for the Fe dimer, the quantum-mechanical excitation energy
is $3 J - 5 |D|$. With $|D|=\unit[1.15\pm0.02]{meV}$, obtained
from the first excitation, this gives an exchange constant $J$ of $\unit[16\pm 1]{meV}$.
The extremely short lifetimes of the noncollinear state
is due to the exchange interaction. This interaction conserves $S_{z}$
but may alter $S$, so that a noncollinear state can rapidly decay into
a collinear excited state with identical $S_{z}$. To elucidate
this experimental finding we estimated the exchange
energy of a Fe dimer on Pt(111) from first-principles calculations.
The least-energy configuration is a dimer with Fe atoms occupying
next-nearest neighbor fcc and hcp positions. The magnetic moments of
$\unit[2.85]{\mu_{\mathrm{B}}}$ per atom align parallelly and the exchange
constant $J$ is estimated to $\unit[11]{meV}$, in agreement with
experiment.

The classical MAE found for individual Fe and Co
adatoms and clusters predicts stable magnetic configurations on the
time scale of years at temperatures below $\unit[2]{K}$. When treating
the cluster as a quantum-mechanical system, tunneling between magnetic
states has to be considered \cite{Bertaina08}. The magnetization tunneling is
induced either by a transversal magnetic field (absent in our studies)
or by an in-plane anisotropy. The threefold-rotational symmetry of the
substrate results in MAE terms of at least sixth order which
contribute only for $S \ge 3$ \cite{footnote2} and can thus be
excluded for Fe and Co.  An additional mechanism for spin reversal is,
however, provided by the electron-electron interaction.  Hybridization
of the adatom states with those of the Pt substrate leads to short
lifetimes and increased energy smearing of the excited state.  Thus,
an overlap of the ground state and the excited state is likely despite
the large MAE's.  It results in a nonzero transition probability
between the degenerate ground states via an intermediate excited state.

In conclusion, the magnetic stability of atoms and small clusters is
related to the hybridization in two opposing ways. On one hand, the
hybridization increases the anisotropy, stabilizing the spin.  On the
other it decreases the lifetime of the excited state, hence
destabilizing the spin. Only by decoupling these effects, the aim of
magnetically stable atoms and clusters may be reached. The presented
ITS approach will help to find the best combination of magnetic atoms
and substrates to achieve this goal.

The authors acknowledge funding by the Japan Society for the Promotion 
of Science (T.M.) and the Deutsche Forschungsgemeinschaft (W.W.).


\begin{thebibliography}{25}
\bibitem[{foo({\natexlab{a}})}]{footnote}
\bibinfo{note}{By considering the impurity as a quantum spin with $S = n / 2$
  (integer $n$ ), hybridization with the substrate is neglected. Treating the
  hybridization would require an approach which accounts for the itinerant
  character of the impurity.}
\bibitem[{foo({\natexlab{b}})}]{footnote2}
\bibinfo{note}{Note that a sixfold in-plane anisotropy is proportional to
  $(\hat{S}_{+}^{6} + \hat{S}_{-}^{6})$ and thus vanishes if $|S| < 3$.
  $\hat{S}_{+}$ and $\hat{S}_{-}$ are the spin `ladder' operators.}
\end{thebibliography}
\end{document}